# Disparate Patterns of Movements and Visits to Points of Interests Located in Urban Hotspots across U.S. Metropolitan Cities during COVID-19

Qingchun Li[1], Liam Bessell[2], Xin Xiao[3], Chao Fan[4], Xinyu Gao[5], Ali Mostafavi[6]

[1] Ph.D. student, Zachry Department of Civil and Environmental Engineering, Texas A&M University, 199 Spence St., College Station, TX 77843-3136; e-mail: qingchunlea@tamu.edu
[2] Undergraduate student, Department of Computer Science and Engineering, Texas A&M University, 199 Spence St., College Station, TX 77843-3112; e-mail: lbessell@tamu.edu
[3] Undergraduate student, Department of Computer Science and Engineering, Texas A&M University, 199 Spence St., College Station, TX 77843; e-mail: xyx56@tamu.edu
[4] Ph.D. student, Zachry Department of Civil and Environmental Engineering, Texas A&M University, 199 Spence St., College Station, TX 77843-3136; e-mail: chfan@tamu.edu
[5] Ph.D. student, Zachry Department of Civil and Environmental Engineering, Texas A&M University, 199 Spence St., College Station, TX 77843-3136; e-mail: xy.gao@tamu.edu
[6] Assistant Professor, Zachry Department of Civil and Environmental Engineering, Texas A&M University, 199 Spence St., College Station, TX 77843-3136; e-mail: amostafavi@civil.tamu.edu

**ABSTRACT**

We examined the effect of social distancing on changes in visits to urban hotspot points of interest. The understanding of population movements and mobility is critical to the modeling of and subsequent containment of pandemics. Urban hotspots, such as central business districts, are gravity activity centers orchestrating movement and mobility patterns in cities. In a pandemic situation, urban hotspots could be potential superspreader areas as visits to urban hotspots can increase the risk of contact and transmission of a disease among a population. In this study, we mapped origin-destination networks from census block groups to points of interest (POIs), such as restaurants, museums, and schools, in sixteen cities in the United States. We adopted a coarse-grain approach to cluster origin and destination nodes into hotspots and non-hotspots and studied movement patterns of visits to POIs among the clusters from January to May 2020. Also, we conducted chi-square tests to identify POIs with significant flux-in changes during the analysis period. The results showed disparate patterns across different cities in terms of reduction in POI visits to hotspot areas. The sixteen cities are divided into two categories based upon movement patterns related to visits to POIs in hotspot areas. In one category, which includes the cities of, San Francisco, Seattle, and Chicago, we observe a considerable decrease in visits to POIs in hotspot areas, while in another category, comprising the cites of, Austin, Houston, and San Diego, among others, the visits to hotspot areas did not greatly decrease during the social distancing period. In addition, while all the cities exhibited overall decreasing visits to POIs, one category maintained the proportion of visits to POIs in the hotspots. The proportion of visits to some POIs (e.g., Restaurant and Other Eating Places) remained stable during the social distancing period, while some POIs had an increased proportion of visits (e.g., Grocery Stores). We also identified POIs with significant flux-in changes, which indicated that related businesses were greatly affected by social distancing measures. The findings highlight that social distancing orders do yield disparate patterns of reduction in movements and visits to POIs in urban hotspots. The reduction of visits to POIs in urban hotspots is an important component of containing pandemics and epidemics. The findings could provide insights for better evolution and monitoring of the effectiveness of social distancing measures in reducing visits to POIs in hotspot areas.

**INTRODUCTION**

The objective of this study is to examine movement patterns to urban hotspots in United States cities during the initial 2020 COVID-19 outbreak. Urban mobility and movement patterns are important characteristics of urban dynamics, reflecting the collective human behavior and social interactions (Hamedmoghadam et al. 2019). Urban mobility drives the spatial flux of populations, and effective epidemic control measures greatly rely on characterization of urban mobility patterns (Danon et al. 2009; Le Menach et al. 2011; Merler and Ajelli 2010; Wesolowski et al. 2012). Assessment of urban mobility is a critical element of epidemic control (Gao et al. 2020). Most standard epidemic models employ mobility patterns in prediction of a disease outbreak trajectory. Tizzoni et al. (2014) used commuter movement data to model the spatial spread of epidemics in European countries. The study examined whether the mobility data matched the empirical mobility pattern and how the observed discrepancies of mobility patterns would affect the results of influenza-like illnesses spread simulation. Balcan et al. (2009) developed a worldwide epidemic model to evaluate the force of infection based on the description of mobility patterns obtained by the gravity model. The results showed that long-range airline traffic determined the global epidemic dynamic, while the short-range mobility patterns determined the local epidemic diffusion pattern. Ferguson et al. (2005) developed a transmission model for the H5N1 influenza in Southeast Asia taking the community mobility into consideration. The model tested containment strategies such as prophylaxis and social distancing measures under different reproduction number of the virus. Meloni et al. (2011) found that it is essential to consider how the epidemic directives enacted by states, for instance, induce changes in mobility patterns and how the changes in turn affect the propagation of the epidemic. Meloni et al. (2011) developed an epidemic model taking into consideration changes of mobility patterns due to the response to an epidemic outbreak. The results showed that self-initiated behavioral changes (e.g., changes in traveling routes) may accelerate the spread. These studies and models highlight the necessity of characterizing mobility and movement patterns for better understanding the extent and trajectories of COVID-19 in metropolitan urban areas.

While the reduction in overall movements and mobility could promote containment, it is equally important to monitor and evaluate movement reduction to urban hotspots. In comparing the effectiveness of social distancing measures between cities, it has been observed that epidemic spread trajectories are different, while the overall mobility reduction is similar across cities. These disparate trajectories could be in part due to differences in movement patterns to urban hotspots. Urban hotspots and sub-centers usually have higher populations and employment densities and more points of interest (POIs) compared with other areas of cities (McMillen 2001; McMillen and Smith 2003). Urban hotspots and sub-centers, therefore, are gravity activity centers affecting population movement, mobility patterns, and human interactions. In a pandemic situation, however, urban hotspots could be potential 'superspreader' POIs (Chang et al. 2020), because visits to hotspots can greatly increase the risk of contact and transmission of a disease. Understanding mobility patterns of the visiting of urban hotspots is critical for developing and monitoring effective epidemic control measures. Origin-destination (OD) networks, under such a situation, are especially helpful for locating hotspots and for studying the urban mobility patterns of visiting urban hotspots (Chang and Tao 1999; Egu and Bonnel 2020; Oliver et al. 2020). Louail et al. (2014, 2015) and Hamedmoghadam et al. (2019) used the OD matrix and a coarse-grain approach to study the mobility among hotspots and non-hotspots. The OD matrices aggregate mobility of individuals from one point to another (Ortúzar and Willumsen 2011; Weiner 2016). Therefore, the OD matrices include insightful information of population movements and patterns of movements within and across cities (Louail et al. 2015; Mazzoli et al. 2019). In addition to traditional surveys and counting to develop OD matrices, increasing studies extracted OD matrices based on the emerging digital footprint data (Barbosa et al. 2018; Blondel et al. 2015). Mazzoli et al. (2019) extracted the OD matrices from Twitter data to map daily commuting flows in London and Paris. Lenormand et al. (2014) mapped the OD matrices from three datasets, including Twitter, mobile phone, and census data. This study showed strong correlations between three datasets regarding individual mobility patterns, lending support to interchanging the three datasets to study mobility patterns.

In this paper, we discuss movement patterns related to a population's visits to urban hotspots using origin-destination networks from census block groups (CBGs) to points of interest (POIs) in 16 cities of United States based on the digital trace data from SafeGraph. The POI data enable identification of urban hotspots to enable evaluation of movement patterns and changes in visiting urban hotspots due to social distancing measures during COVID-19. We adopted a modified coarse-grain approach to clustering the hotspots and non-hotspots nodes in OD networks (Hamedmoghadam et al. 2019). Then, we studied movement patterns related to visiting POIs among hotspots and non-hotspots across 16 cities. Figure 1 illustrates the conceptual model of four types movements among hotspots and non-hotspots. Furthermore, we conducted chi-square tests to identify what POIs had significant flux-in changes during COVID-19. As each POI node is associated with a business category, the results could also help identify what business industries were affected due to social distancing during COVID-19.

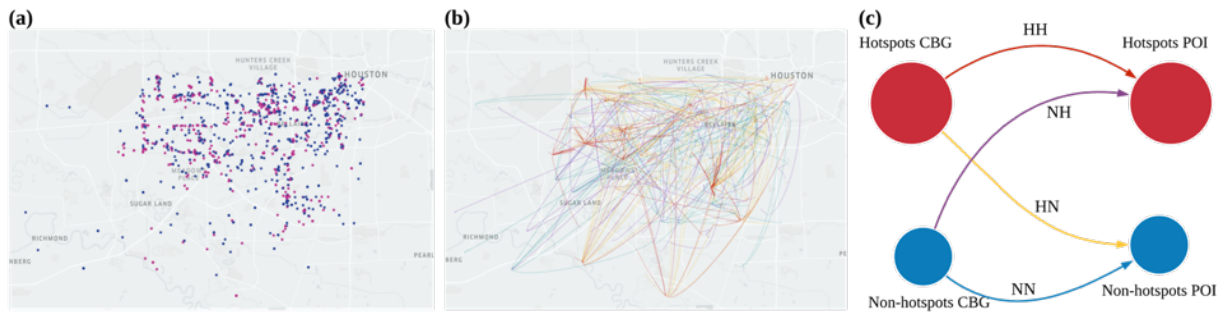

**Figure 1**. Coarse-grain approach to categorize origin-destination movements to four types of movements among hotspots and non-hotspots: (a) hotspots and non-hotspots, (b) individual OD movements, (c) conceptual representation of four types of movements., (Figure is based on Houston SafeGraph data.)

**DATA AND METHODOLOGY**

We used point of interest data provided by SafeGraph to map the origin-destination network. SafeGraph aggregates POI data from diverse sources (e.g., third-party data partners, such as mobile application developers), and removes private identity information to anonymize the data. The POI data include base information of a POI, such as the location name, address, latitude, longitude, brand, and business category. SafeGraph uses standard North American Industry Classification System (NAICS) to classify POI business categories. The data reveal the visit pattern of POIs including the aggregated number of visits to the POI during the data range, the number of visits to the POI each day over the period, and the aggregated number of visitors to the POI from census block groups during the period (e.g., one week and one month).

In this paper, we used the POI data: Weekly Pattern Version 2, to study movement patterns across sixteen cities in the United States (SafeGraph 2020). Among these 16 cities are 14 largest cities in United States by population. In addition, Seattle and Detroit were studied. Seattle was the first city in the United States to report a diagnosed COVID case, and Detroit had a burst in the number of cases in March 2020. The analysis comprises four major steps: 1) map the OD network, 2) Identify hotspots and non-hotspots based on the mapped OD network, 3) examine movement patterns between hotspots and non-hotspots, and 4) identify POIs with significant flux-in changes. We explain each step in the following sections.

### *1. Map the OD network*

We mapped the OD movement network based on the number of visitors to POIs from CBGs. The mapped OD networks are directed and weighted bipartite networks. The partite of origin nodes are CBGs and the partite of destination nodes are POIs. Links in the OD network represent visits from CBGs to POIs, and weights of links are the number of visitors during the covered period. We mapped the weekly OD network because SafeGraph aggregates the number of visitors from CBGs to POIs by week. Figure 2 illustrates an example of the mapped OD network in Jacksonville, Florida.

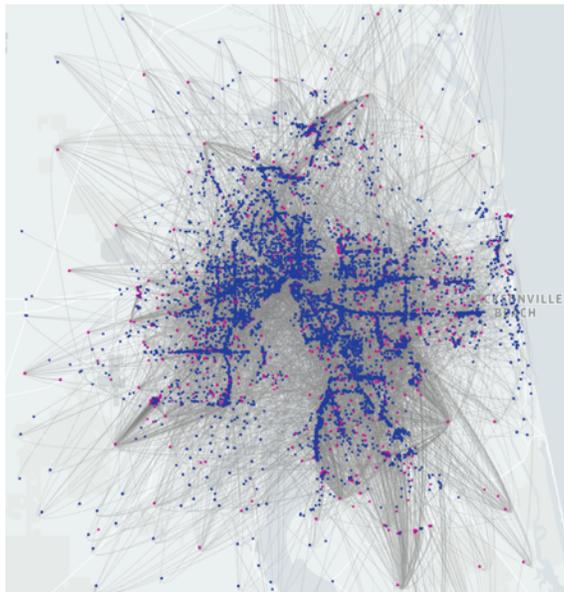

**Figure 2.** Mapped OD network of the week of January 27, 2020, in Jacksonville, Florida. Total 83,661 visits. Red nodes are hotspots (1,314 nodes) and blue nodes are non-hotspots (10,820 nodes)

### *2. Identify hotspots and non-hotspots*

In the literature, different methods have been proposed to separate hotspots and non-hotspots (Giuliano and Small 1991; McMillen 2001; McMillen and Smith 2003). Louail et al. (2014, 2015) developed a method to identify hotspots and non-hotspots based on the Lorenz curve of divided $1-km^2$ cells. This method yields lower and upper boundaries of identified hotspots. In this paper, we adopted a centroid-based clustering method to separate hotspot and non-hotspot nodes in the mapped OD network (Hamedmoghadam et al. 2019). Each mapped weekly OD network has a correspondent OD bi-adjacency matrix. The columns and rows of the OD matrix represent origin nodes and destination nodes, and the elements are the weights of links. First, we summed all the rows and columns to get the total flux-out and flux-in values of origin and destination nodes, respectively. Then, we sorted flux-out and flux-in values of origin and destination nodes in an ascending order: $O_1 < O_2 < \cdots < O_n$ and $D_1 < D_2 < \cdots < D_n$. To separate hotspots and non-hotspots in these two lists, we used Equation 1 to determine the separation point $O_c$ and $D_c$. Nodes with flux-out and flux-in values greater than $O_c$ and $D_c$ are hotspots of origins and destinations. In Equation 1, $q_i$ could represent either $O_1, O_2, \cdots, O_n$ or $D_1, D_2, \cdots, D_n$.

$$\arg\min_c \sum_{i=1}^{c} |q_i - \frac{1}{c}\left(\sum_{k=1}^{c} q_k\right)| + \sum_{j=c+1}^{n} |q_j - \frac{1}{n-c}\left(\sum_{l=c+1}^{n} q_l\right)| \qquad (1)$$

### *3. Examine movement patterns between hotspots and non-hotspots*

We used a coarse-grain approach to examine the mobility pattern (Hamedmoghadam et al. 2019; Louail et al. 2014, 2015). The approach reduces the mobility flows to four types: (1) $HH$: from hotspot origins (CBGs) to hotspot destinations (POIs), (2) $NH$: from non-hotspot origins to hotspot destinations, (3) $HN$: from hotspot origins to non-hotspot destinations, and (4) $NN$: from non-hotspot origins to non-hotspot destinations. If we use $F$ to represent the original OD bi-adjacency matrix, then we could reduce $F$ to the coarse-grained matrix $\Lambda$.

$$\Lambda = \begin{bmatrix} HH & NH \\ HN & NN \end{bmatrix} \qquad (2)$$

In matrix $\Lambda$, each sub-matrix could be calculated as follows pattern (Hamedmoghadam et al. 2019; Louail et al. 2014, 2015).

$$HH = \sum_{i \in M, j \in p} F_{ij} / \sum_{i,j} F_{ij} \qquad (3)$$

$$NH = \sum_{i \notin M, j \in p} F_{ij} / \sum_{i,j} F_{ij} \qquad (4)$$

$$HN = \sum_{i \in M, j \notin p} F_{ij} / \sum_{i,j} F_{ij} \qquad (5)$$

$$NN = \sum_{i \notin M, j \notin p} F_{ij} / \sum_{i,j} F_{ij} \qquad (6)$$

where $F_{ij}$ represents each element in the original OD matrix, $M$ represents the set of hotspot origins, and $P$ represents the set of hotspot destinations determined in step 2. Equations 3 through 6 illustrates how we calculated the proportion of each types of movements. We normalized each mobility type by the total mobility flow. Therefore, the proportion of each type of movement, $HH$, $NH$, $HN$, and $NN \in [0,1]$ and the sum of them equals 1. $HH$, $NH$, $HN$, and $NN$ could represent the proportion of each type of movement flow in the whole OD network (Hamedmoghadam et al. 2019).. This characterization is particularly important to examine and monitor reduction in movements to urban hotspots (reduction in the proportion on *HH* and *NH* movements) during social distancing periods.

### *4. Clustering analysis of movement patterns across cities*

We conducted clustering analysis after we obtained movement patterns of cities. We used the sum of proportions of two movements, $HH + NH$, as an indicator of cluster movement patterns within cities. These two movements would contribute to spreading of the epidemic because the extent of visits to the hotspot POIs could increase the transmission rate of COVID-19. We scaled the time series data related to movement patterns so that each time series has zero mean and unit standard deviation. This step enabled us to focus on comparing the shapes and trends of time-series data. We compared three algorithms (Euclidean distances, dynamic time warping (DTW), cross correlation) for time series clustering (Cuturi and Blondel 2017; Paparrizos and Gravano 2015; Petitjean et al. 2011) and used the silhouette coefficient to determine the number of clusters (Rousseeuw 1987). (Results of algorithms are within the supplemental information.)

### *5. Identify POIs with significant flux-in changes*

We compared OD matrices from two milestone dates (e.g., March 1 and March 29). We summed columns of the matrices to obtain the weighted node degree centrality of destinations. Then we calculated differences in the weighted degree centrality of each pair of destination nodes in the two matrices: $C_1, C_2, \cdots, C_n$. Accordingly, $C_n^2 / \overline{C^2}$ approximately follows the chi-square distribution if the weighted node degree centrality did not have significant changes. (Proof process is discussed in the supplemental information.) Here, $\overline{C^2}$ is the average of the square of degree centrality difference. We used the upper tail test ($H_1: |C| > 0, H_0: |C| = 0$) of the chi-square distribution to determine the P-value for each node. Because each test was conducted separately for each node, the degree of freedom is 1, and we adjusted P-value for multiple tests using the Benjamini-Hochberg (B-H) false discovery rate (FDR) correction (Benjamini and Hochberg 1995). We tested the destination node set and identified the POIs with significant in-degree changes (with FDR equal to 0.1 and adjusted P-value < 0.01), which could reflect significant flux-in changes. Furthermore, because each POI has its NAICS code indicating its business activity, we can identify the extent to which business activities were impacted due to COVID-19.

**RESULTS**

### *1. Movement patterns of visiting POIs in 16 cities*

The first set of results shows that the sum of absolute visits to POIs showed a decreasing trend for all 16 cities after the enforcement of shelter-in-place orders (Figure 3). However, four types of movements (*HH, HN, HH,* and *NH*) varied across different cities. Figure 3 includes the result of clustering analysis, and the 16 cities were divided into two categories. Figure 5 illustrates the detailed clustering results.

We can observe from Figure 3 that the *HH* movements in category 1 cities (San Diego, Fort Worth, Dallas, Houston, and Austin) did not show a clear declining trend. In category 1 cities, a decline in *HN* and *NN* movements caused a decrease in the total number of visits. In fact, the *NH* movements in cities of category 1 (except for New York) showed an increasing trend after the enforcement of shelter-in-place order. In category 2 cities, *HN* and *NN* movements remained stable, while *HH* and *NH* movements show a clear downward trend.

We also investigated the proportion of each type of movement in 16 cities. We can observe from Figure 4 that the proportion of *HH* and *NH* movements in category 1 cities did not show a declining trend, even though the absolute value of *HH* and *NH* movements in some cities, such as Phoenix, San Antonio and New York, declined (as shown in Figure 3). The proportion of *HN* and *NN* movements in most cities (except for New York) of category 1 did not show a clear upward trend. This result demonstrates that although people in category 1 cities had decreased their absolute visits to POIs, the proportion of their visits to the hotspots of POIs were stable.

For category 2 cities, the proportion of *HH* and *NH* movements showed a clear downward trend, while the proportion of *HN* and *NN* movements had a clear upward trend. The barycenter of two city categories illustrated in Figure 5 indicates that the proportion of *HH* and *NH* movements had an upward trend in category 1 cities, while the proportion of *HH* and *NH* movements showed a downward trend in cities of category 2. These results imply that people in cities of category 2 kept reducing the proportion of their movements to urban hotspots of POIs due to the social distancing measure. We can conclude from the above results that while the overall mobility in all cities declined due to social distancing orders, the movement patterns related to visits to hotspots followed two different trends in the two categories of studied cities. The disparate patterns could imply differences in transmission risks.

In addition, we can observe from Figure 4 that the proportion of *HH* and *NH* movements in some category 2 cities (Detroit, Jacksonville, Chicago, Los Angeles, San Francisco, Seattle, and San Jose) started to decline much earlier than the enforcement of the shelter-in-place orders. This result may imply that people in these cities had started to proactively reduce visits to hotpots. The first cases occurred quite early in most of these cities, such as Detroit, Chicago, Los Angeles, Seattle and San Jose. As there is a clear gap between the date of the first case and the enforcement of shelter-in-place orders in these cities, this result may suggest that the information of the first case may trigger proactive actions.

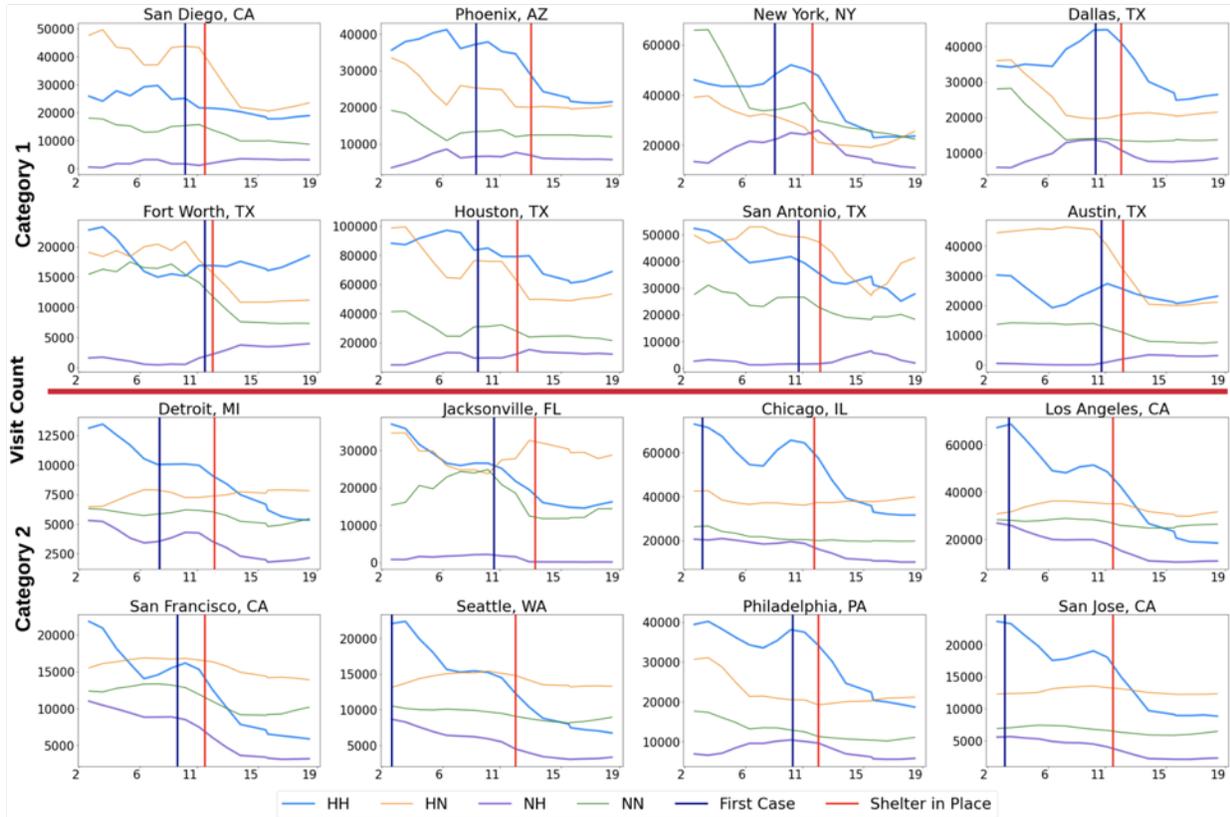

**Figure 3**. Absolute visits related to four types of movements in 16 cities (weeks from December 30, 2019 to May 11, 2020). We used the rolling mean (window = 4) to smooth the data, original data could refer to supplemental document.

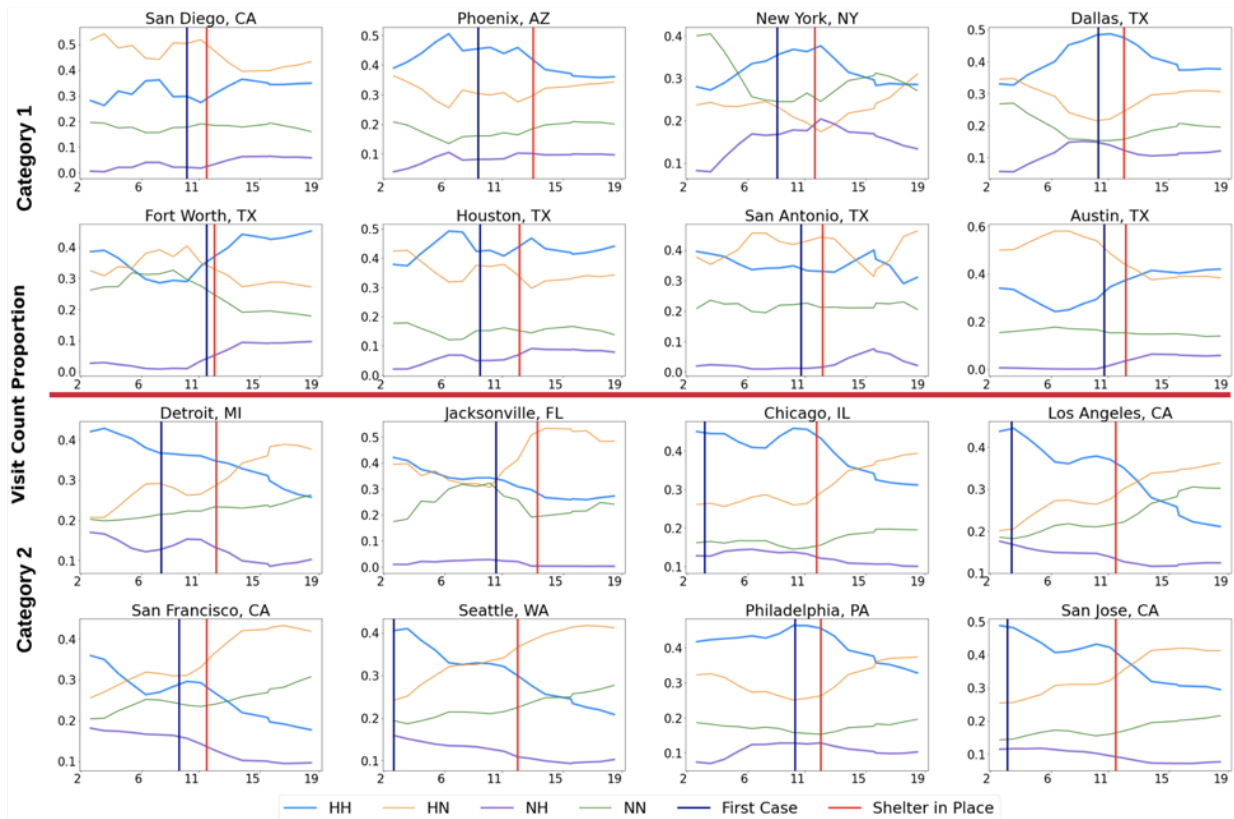

**Figure 4.** Proportion of four types of movements in 16 cities (weeks from December 30, 2019 through May 11, 2020), rolling mean (window = 4)

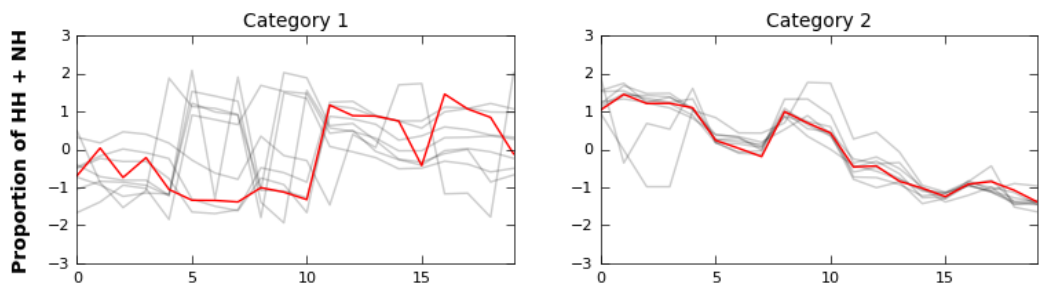

**Figure 5.** Result of clustering analysis using dynamic time warping barycenter averaging (weeks from December 30, 2019 through May 11, 2020). Each gray line represents movement of one city in categories; the red line represents the barycenter of the category

## 2. Proportion of persons visiting POIs in hotspots

The next set of results indicates one dominant POI in hotspots across all 16 cities: Restaurants and Other Eating Places (NAICS code: 7225). Museums, Historical Sites and Similar Institutions (NAICS code: 7121) was the second dominant POI in many of the studied cities. Based on the description of NAICS, Museums, Historical Sites and Similar Institutions encompasses several sub-categories, including Museums, Historical Sites, Zoos and Botanical Gardens, and Nature Parks and Other Similar Institutions. Surprisingly, the proportion of visits to these two POIs remained fairly stable during the unfolding of the COVID-19 pandemic and the enforcement of shelter-in-place orders. Figure 6 illustrates the top seven hotspot POIs with the highest proportion of visits in four cities: Austin, New York, San Francisco, and Seattle (The results for the other cities are provided in the supplemental information.) We selected two cities from each category and one week at the end of each January, February, March, and April to illustrate the patterns.

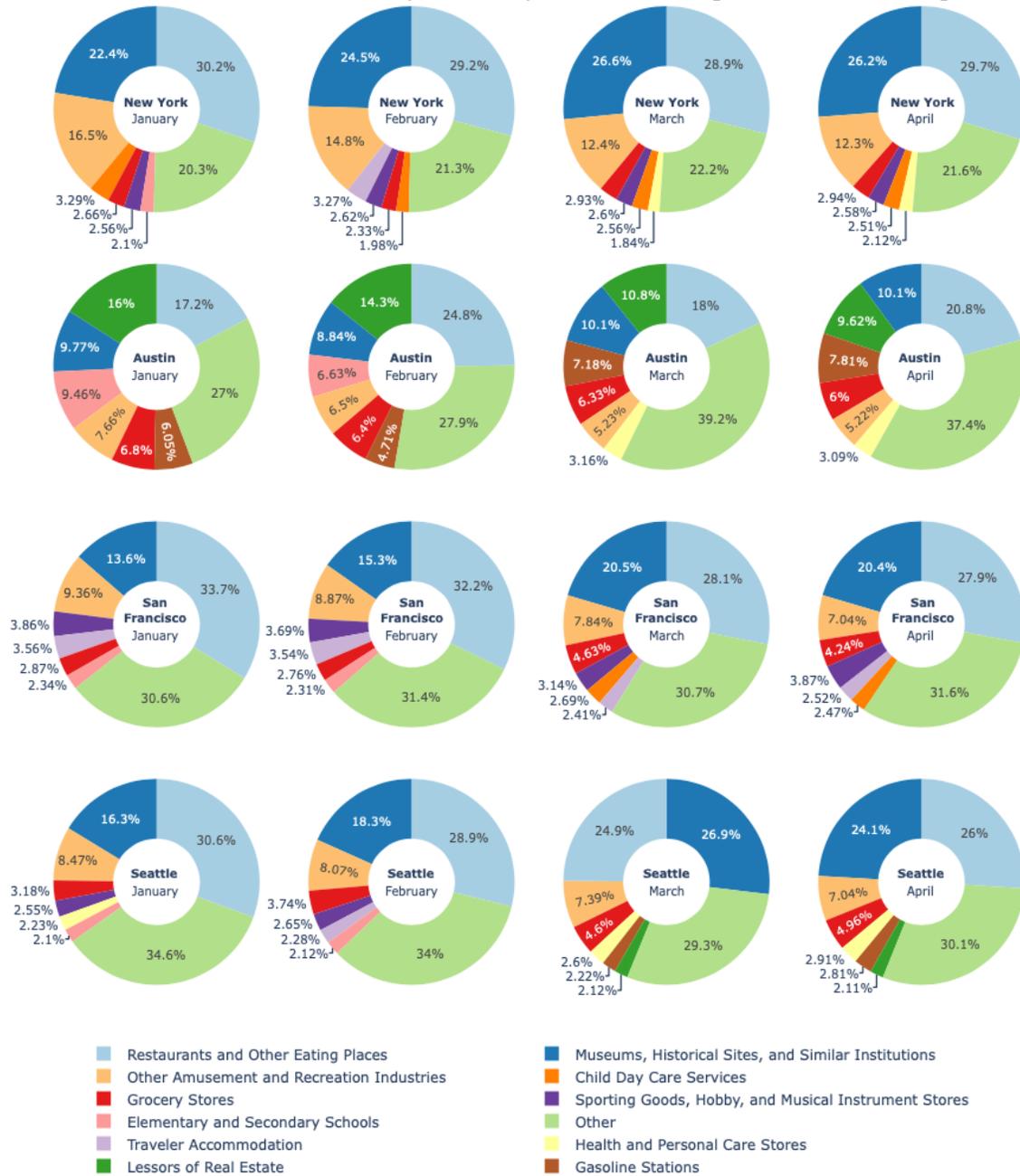

**Figure 6.** Top 7 hotspot POIs in hotspots during the last week of January, February, March, and April.

As illustrated in Figure 6, the proportion of visits to POIs in hotspots showed a similar pattern in the weeks of January 27, 2020, and February 24, 2020, in addition to the two dominant POIs (i.e., restaurants and museums), Other Amusement and Recreation Industries (NAICS code: 7139) ranked third for three cities (ranked fourth in Austin), while the fourth and fifth place POIs varied across cities: Child Day Care Services (NAICS code: 6244) and Traveler Accommodation (NAICS code: 7211) in New York; Sporting Goods, Hobby, and Musical Instrument Stores (NAICS code 4511) and Traveler Accommodation (NAICS code: 7211) in San Francisco; as well as Sporting Goods, Hobby, and Musical Instrument Stores (NAICS code: 4511) and Grocery Stores (NAICS code: 4451) in Seattle. In Austin, Lessors of Real Estate (NAICS code: 5311), Gasoline Stations (NAICS code: 4471), and Elementary and Secondary Schools (NAICS code: 6111) had large proportions of visits. With the unfolding of COVID-19 and shelter-in-place orders, although the proportion of visits to the top two POIs slightly decreased, the top two POIs in each city still were the dominant places visited. After the unfolding of COVID-19 and social distancing orders, the proportion of visits to Grocery Stores (the red element in Figure 6) increased. Also, the proportion of visits to Other Amusement and Recreation Industries and Travel Accommodation declined. For the week of March 30, Grocery Stores started to rank fifth while another essential POI, Gasoline Stations, ranked fourth in Austin. Also, Grocery Stores started to rank fourth in the weeks of March 30 and April 27 in the other three cities. In Austin, the proportion of Grocery Stores visits decreased in the weeks of March 30 and April 27, but the rank increased. In other cities, both the rank and the proportion of Grocery Stores visits increased. We also found that Health and Personal Care Stores (NAICS code: 4461) POIs and General Merchandise Stores, including Warehouse Clubs and Supercenters (NAICS code: 4523) POI showed an upward trend in most of the cities after the outbreak started. For example, the Healthcare POI ranked among the top seven in the weeks of March 30 and April 27 in New York and Austin. In Seattle, the proportion of visits to Health and Personal Care Stores POI rose to fifth place in ranking in the weeks of March 30 and April 27. The proportion of visits to the Merchandise POI rose to the top 7 in the weeks of March 30 and April 27 in Houston, Dallas, Detroit, Phoenix and rose to the top 3 in Jacksonville and San Antonio.

Because we determined POI hotspots based on the total number of visits to POIs, the evolution of the proportion of visits to POIs in hotspots could provide insights about movement patterns of people across different cities. The results showed that although the absolute number of visits decreased for all the POIs during COVID-19, the proportion of visits to restaurants and museums remained dominant in most cities. Also, the results showed that the proportion of visits to grocery stores and healthcare facilities increased, while the proportion of visits to amusement and recreation industries decreased. Furthermore, the patterns of visits to POIs did not show a relationship with city categories based on movements to hotspots. Instead, the visits to POIs highly depended on the attributes of cities. For example, Gasoline Station was the second highest visited POI hotspot in Houston and was third in Dallas and Detroit, while representing only a small proportion of hotspot POI visits in New York. Museums, Historical Sites and Similar Institutions was the second highest proportion of hotspot POI visits in most studied cities, such as Dallas, Detroit, Philadelphia, Los Angeles and San Jose, while it formed a small proportion of hotspot POI visits in Jacksonville, Fort Worth and Houston.

### 3. POIs with significant flux-in changes

Based on the number of nodes with significant flux-in changes, we identified several businesses highly affected by the COVID-19 pandemic, including Restaurants and Other Eating Places, Museums, Historical Sites, and Similar Institutions, Lessors of Real Estate, Elementary and Secondary Schools (NAICS code: 6111), Support Activities for Air Transportation (NAICS code: 4881) and Religious Organizations (NAICS code: 8131). Also, some of the affected POIs varied across the 16 cities. Figure 5 illustrates the POIs with significant flux-in changes in four selected cities: New York, Austin, San Francisco, and Seattle. (The results for other cities could be found in the supplemental information.) We selected one week at the end of each January, February, March and April to compare the trends with the week of January 13 (with the assumption that most businesses had returned to normal schedules and patterns of visits after the winter break).

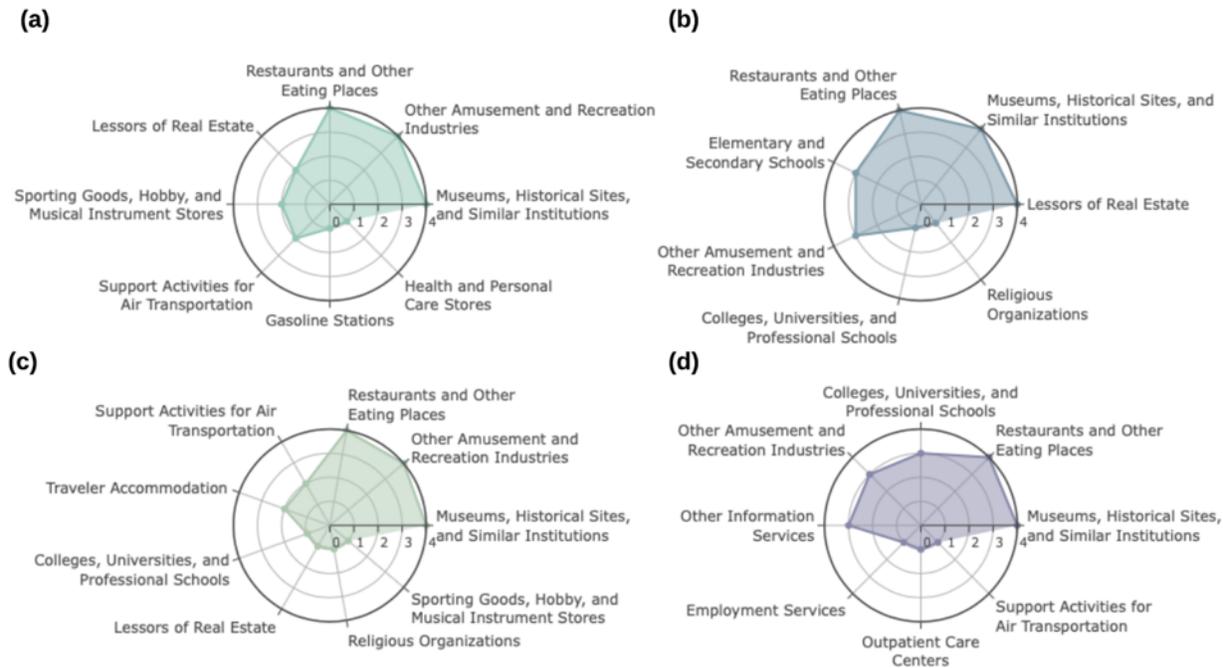

**Figure 5.** POIs with significant flux-in changes in the four one-week periods: (a) New York, (b) Austin, (c) San Francisco, (d) Seattle. We included top 5 businesses related to hotspot POIs with significant flux-in changes in each week. The indicator on the radar chart refers to the number of weeks the business activity was in the top 5 affected POIs.

Figure 5 illustrates that visits to Restaurants and Other Eating Places and Museums and to Historical Sites, and Similar Institution POIs were greatly affected in all four cities. These two POIs ranked in the top 5 affected business activities across all four studied weeks. Other Amusement and Recreation Industries was another highly affected POI, ranking in the top 5 affected POIs four times in New York and San Francisco, and in top 5 affected POIs list three times in Austin and Seattle. Also, the extent of affected POIs varied across different cities, such as Lessors of Real Estate and Elementary and Secondary Schools in Austin, Support Activities for Air Transportation in New York and San Francisco, and College, University and Professional Schools and Other Information Services in Seattle.

The results indicate that some POIs are universally affected across all cities during the January through May time period examined in this study. The effects of the pandemic of other POIs varied across cities and months. For example, the effect on Support Activities for Air Transportation visits were related to travel restrictions which had the greatest impact in New York and San Francisco. We can observe from Figure 4 that Travel Accommodation had relatively large proportion of POI hotspots in New York and San Francisco (ranked top 4 and 5, respectively, before March). Also, the shelter-in-place order affected Elementary and Secondary Schools in Austin and College, University and Professional Schools in Seattle due mainly to closure of schools and colleges.

**DISCUSSION**

The results of this study provide a deeper insight into the effect of social distancing on changes in population visits to hotspot POIs during the COVID-19 pandemic. Although the absolute number of visits to POIs showed a downward trend in the 16 studied cities, one category of cities sustained the proportion of movements to hotspot POIs, while the second category of cities reduced the proportion of movements to hotspot POIs and increased the proportion of movements to non-hotspots POIs. Another COVID-19 study in Italy demonstrated that human mobility in Italy was strongly related to the spread and control of COVID-19 (Cintia 2020). Movements to hotspot and non-hotspot, however, may have different transmission risks and cause different epidemic diffusion patterns. Balcan et al. (2009) considered two types of mobility: long-range mobility and short-range mobility when building an epidemic model. The results showed that two types of mobility determined different epidemic diffusion patterns at regional and local levels. Meloni et al. (2011) showed that changes of mobility patterns due to an epidemic outbreak may have a negative effect on epidemic control. Furthermore, Chang et al. (2020) identified 'superspreader' POIs (e.g., fitness centers and restaurants) that may cause huge amount of infections. Hence, the proportion of movements to POIs in urban hotspots could be a critical indicator of the manner in which cities responded to an epidemic breakout. The results of this study could facilitate better monitoring of the effect of enforced epidemic control measures. Furthermore, we investigated which POIs maintained their pre-epidemic proportion of visits, and which POIs experienced declines and increases in proportion visits during the unfolding of COVID-19 and the enforcement of shelter-in-place orders. The results facilitate a better understanding of human lifestyles and its changes during the epidemic and could inform developing effective epidemic control measures.

Also, we conducted chi-square test to pinpoint POIs with significant flux-in changes. The process could be a good complement to the coarse-grain approach that was adopted to analyze the OD network. The coarse-grain approach clustered nodes to hotspots and non-hotspots and grouped individual OD flows into four types of movements. While the approach could provide a useful picture of human movements among hotspots and non-hotspots, it does not provide information about single POIs. An understanding the flux-in changes for single POIs is important for the examination of pandemics. Because our study focused on the effect of social distancing measures and shelter-in-place orders, the POIs with significant flux-in changes showed a decrease in visits during the studied period. This set of results could provide additional insights regarding community response to COVID-19 and inform the monitoring of the control measure effectiveness. On the other hand, these POIs could expect a significant flux-in increase after the shelter-in-place orders are lifted. Specifying these POIs could provide valuable information to develop reopening policies and strategies (e.g., multi-steps to reopen POIs with significant flux-in changes).

Other research directions could be explored based on the findings of this study. For example, based on the results of the proportions of visits to POIs during the studied period across cities, we could refine the understanding of essential and non-essential services for humans in urban disruptions, such as natural hazards and epidemic outbreaks (Esmalian et al. 2019), and future research could take characteristics of cities into consideration. Furthermore, the results could facilitate the understanding how the urban disruptions would affect business (e.g., what business industries would be more affected during disruption compared with other business), helping to develop business disaster planning and recovery strategies in urban disruptions (Karim 2011; Marshall and Schrank 2014).

The research also has some limitations. We tried to study movement patterns in some less populated cities in United States that were highly affected by COVID-19, such as Randolph,

Terrell, and Early in Georgia, as well as Union, Bergen, and Hudson in New Jersey. The movement data in these cities, however, were very sparse and difficult to build the OD network. The results in this paper, therefore, focused primarily on cities with large populations.

**ACKNOWLEDGEMENT**
The authors would like to acknowledge funding support from the National Science Foundation RAPID project #2026814: Urban Resilience to Health Emergencies: Revealing Latent Epidemic Spread Risks from Population Activity Fluctuations and Collective Sense-making." The authors would also like to acknowledge that SafeGraph provided POI data. Any opinions, findings, and conclusion or recommendations expressed in this research are those of the authors and do not necessarily reflect the view of the funding agency. The authors would like to thank Jan Gerston for copy-editing services.

**SUPPLEMENTAL INFORMATION:**

*1. Results of different clustering algorithms.*

Table S1 illustrates the results of silhouette coefficient of different clustering algorithms and different clustering numbers. Figure S1 illustrates the clustering results of different algorithms. We finally choose DBA (dynamic time warping barycenter averaging) with two clusters.

**Table S1.** Silhouette coefficients of different clustering algorithms and different clustering numbers.

|  Algorithms | Silhouette coefficient | | | | |
| --- | --- | --- | --- | --- | --- |
|  | N=2 | N=3 | N=4 | N=5 | N=6 |
| Euclidean | 0.501 | 0.483 | 0.373 | 0.421 | 0.366 |
| DBA | 0.514 | 0.483 | 0.408 | 0.482 | 0.431 |
| Soft-DTW | 0.501 | 0.483 | 0.408 | 0.451 | 0.407 |
| K-shape | 0.0006 | 0.01150 | -0.067 | 0.1237 | 0.0778 |

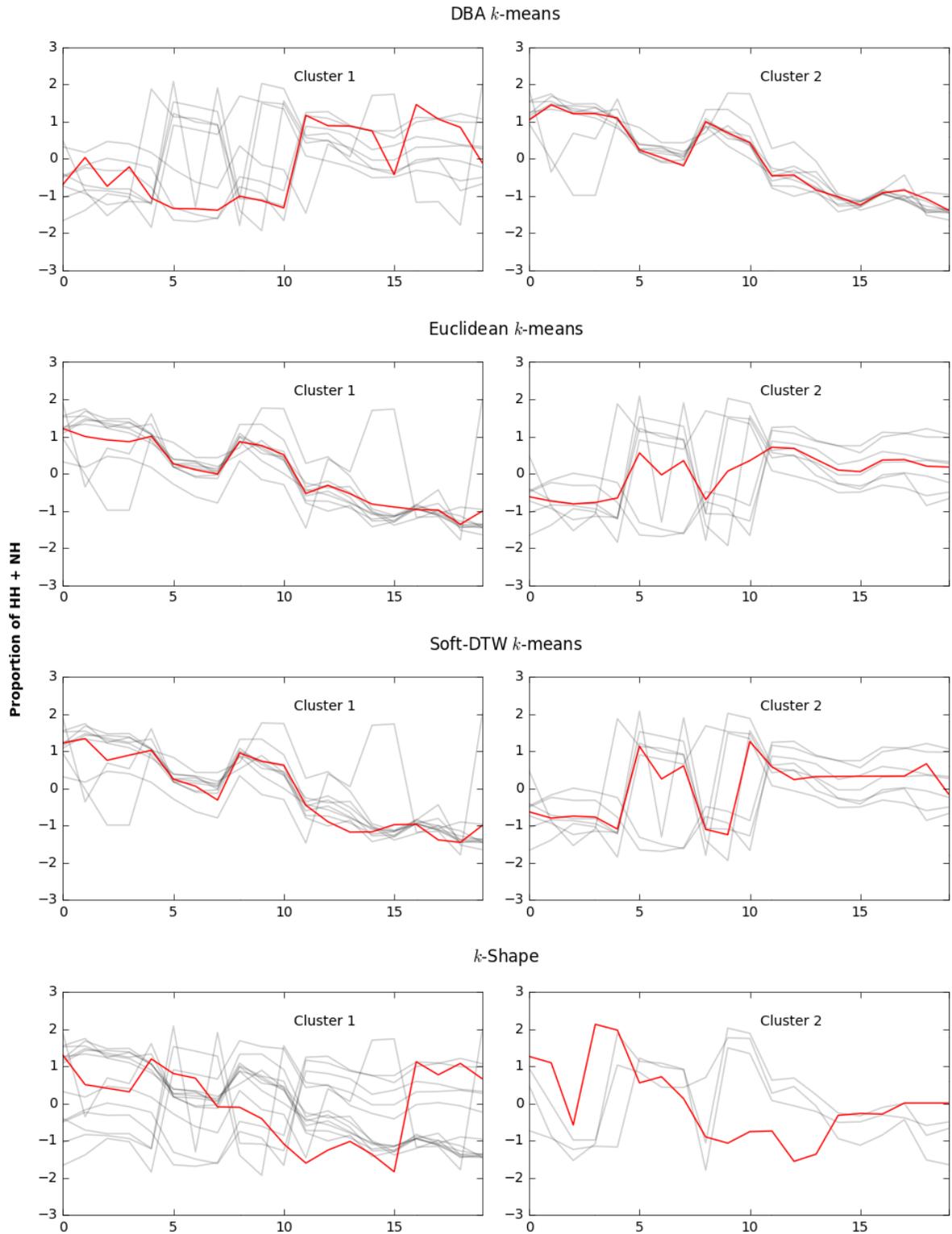

**Figure S1.** Results of clustering algorithms (weeks from December 30, 2019 through May 11, 2020)

## 2. Proof of Chi-square distribution:

For two studied moments (e.g., March 1 and March 29), weighted degree centrality of nodes are $d_{11}, d_{12}, \cdots, d_{1n}$ and $d_{21}, d_{22}, \cdots, d_{2n}$. Then the difference of weighted degree centrality of each pair of nodes are $C_1, C_2, \cdots C_n = (d_{11} - d_{21}), (d_{12} - d_{22}), \cdots, (d_{1n} - d_{2n})$. Therefore, if the weighted degree centrality of nodes in the aggregated weekly OD networks of two studied moments does not have significant changes (null hypothesis), the difference of weighted degree centrality of pair of nodes, $C_1, C_2, \cdots C_n$, approximately follows a normal distribution. The mean of the normal distribution equals to 0, and the standard deviation equals to $\sqrt{\frac{C_1^2 + C_2^2 + \cdots + C_n^2}{n}} = \sqrt{\overline{C^2}}$. Therefore, the $Z^2 = \frac{C_n^2}{\overline{C^2}}$ approximately follows a chi-square distribution with degree of freedom is 1. Osorio et al. (2020) provides the result but does not have the proof process.

## 3. Top 7 POIs in hotspots in cities

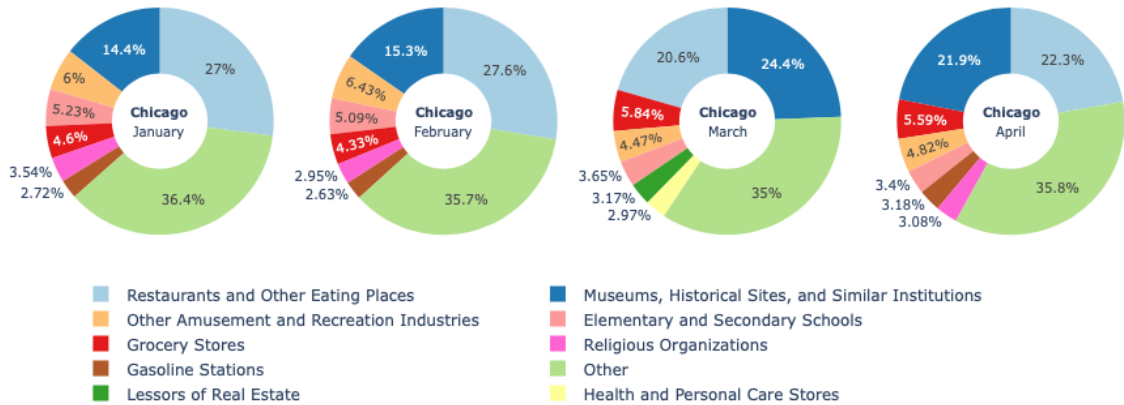

**Figure S2.** Chicago

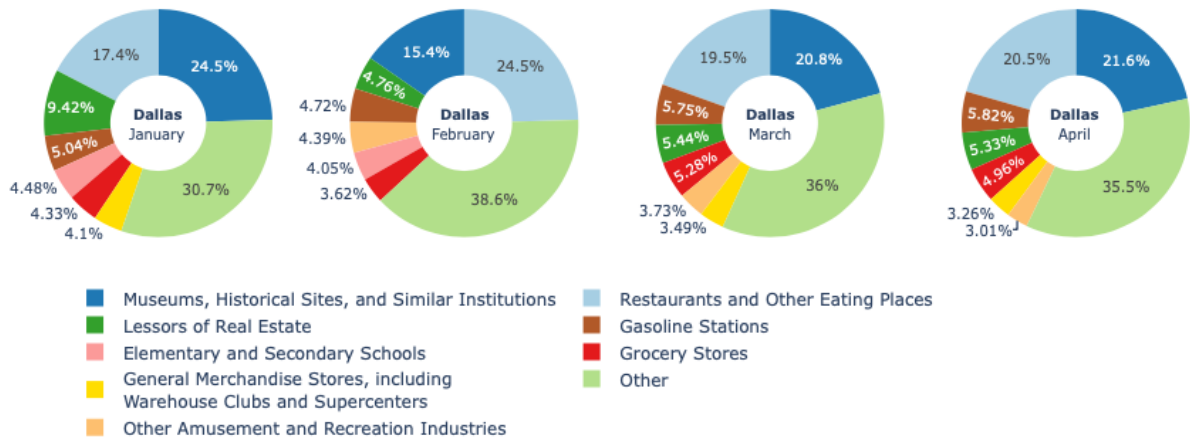

**Figure S3.** Dallas

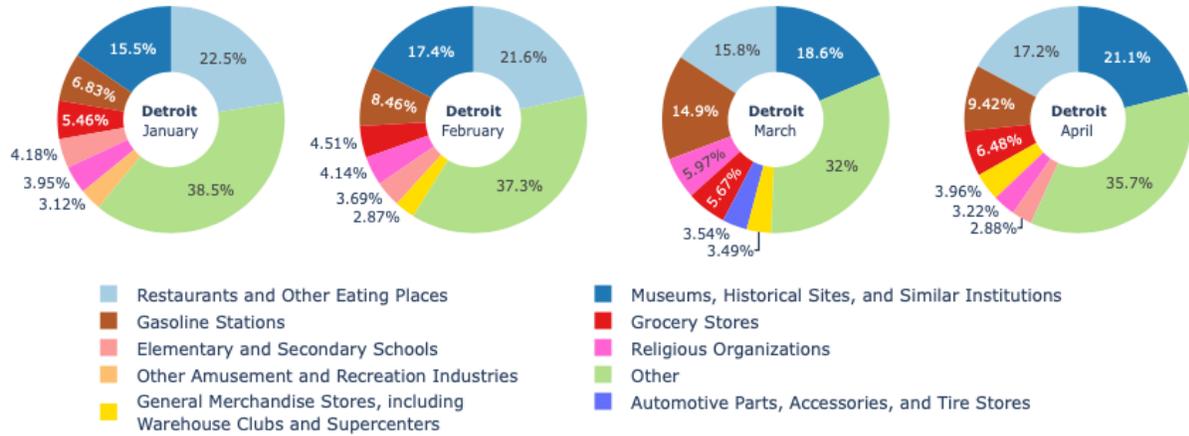

**Figure S4.** Detroit

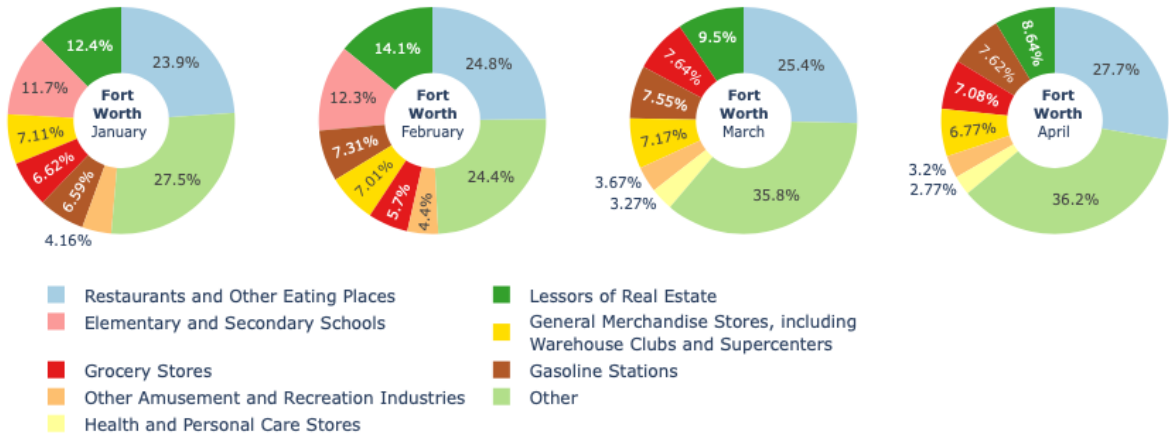

**Figure S5.** Fort Worth

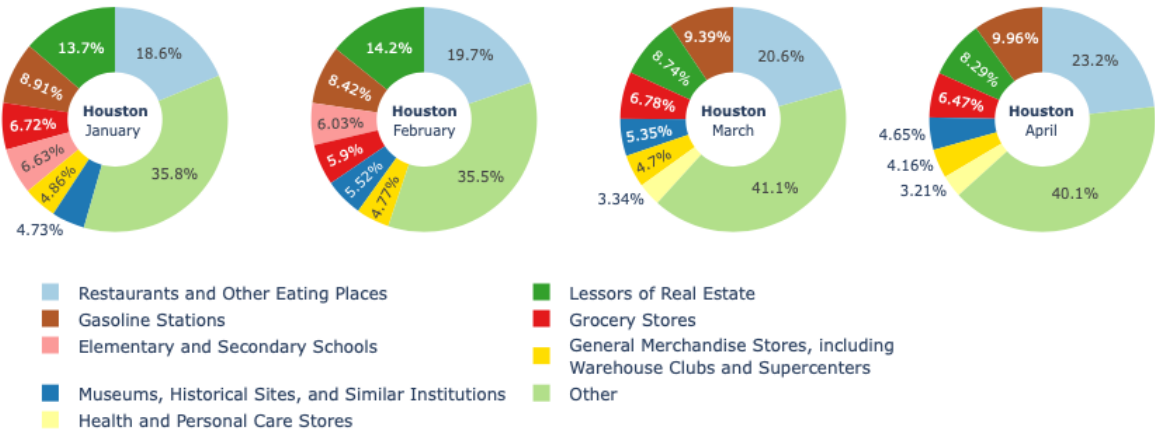

**Figure S6.** Houston

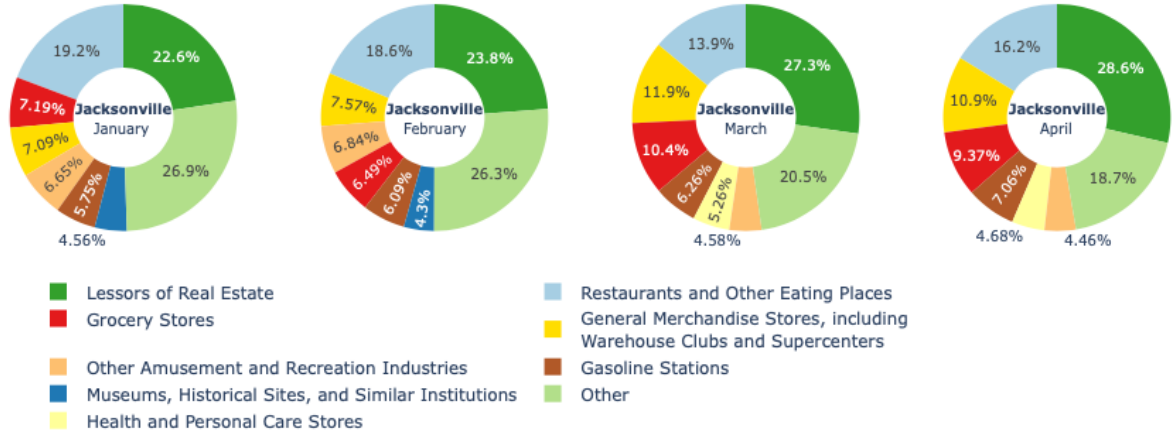

**Figure S7.** Jacksonville.

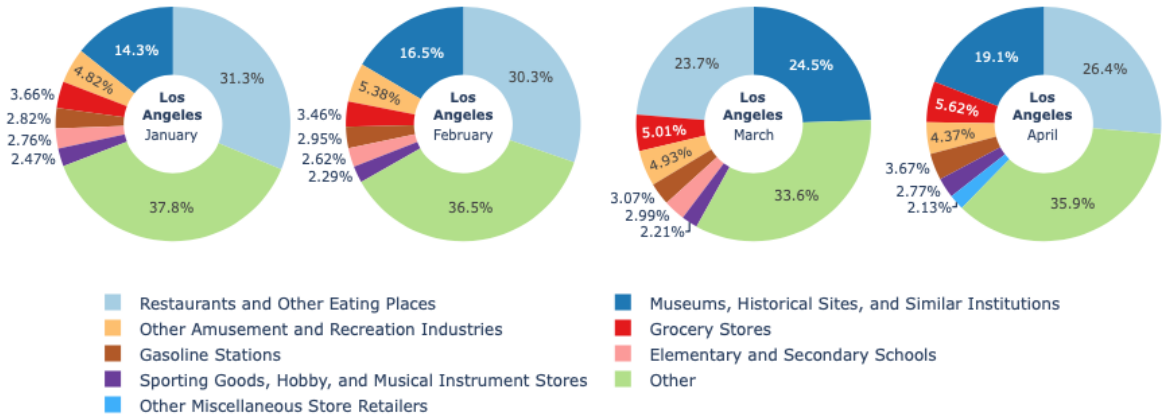

**Figure S8.** Los Angeles.

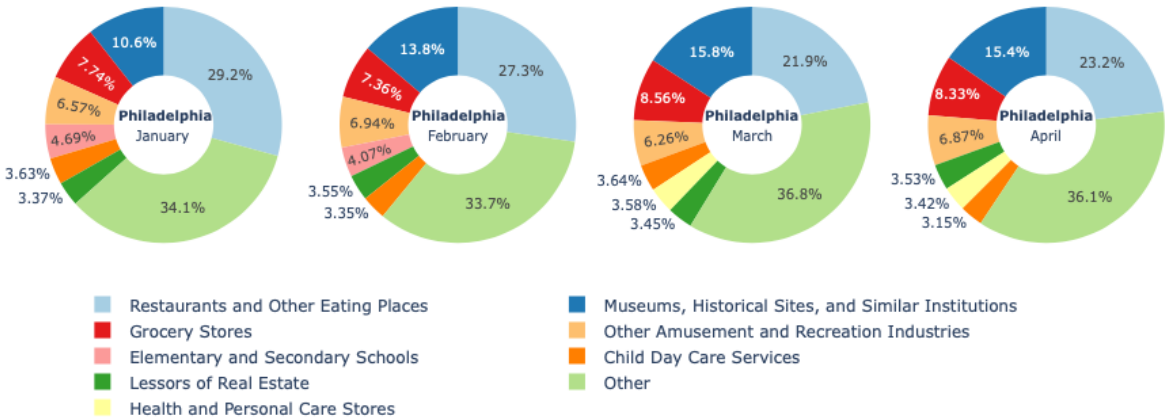

**Figure S9.** Philadelphia.

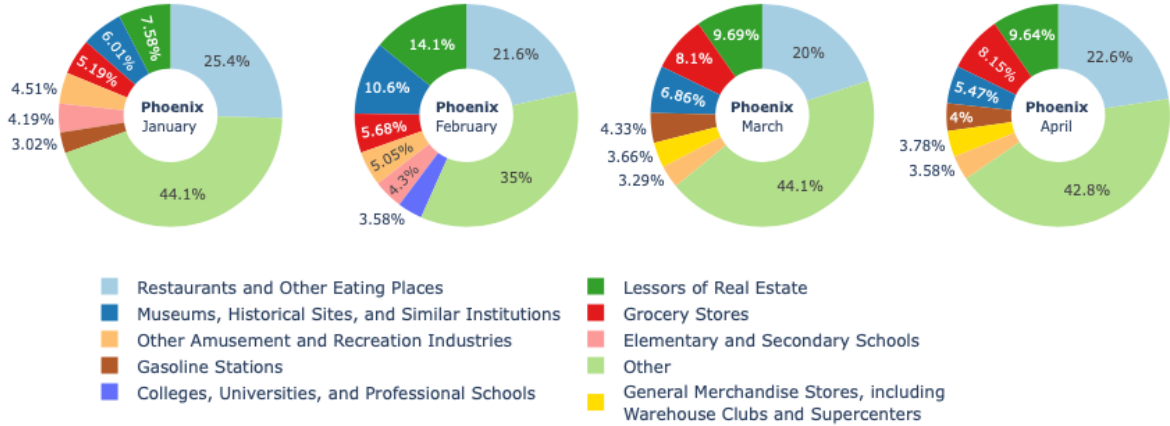

**Figure S10.** Phoenix.

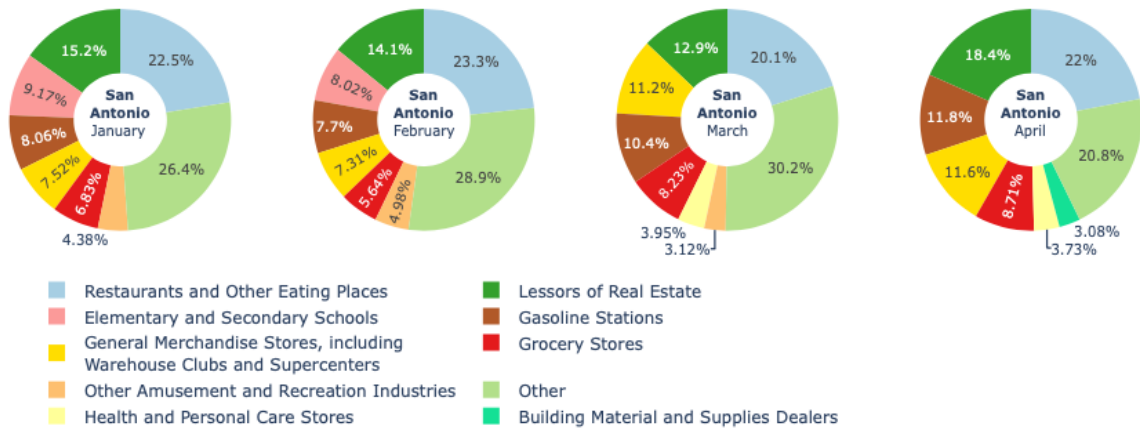

**Figure S11.** San Antonio.

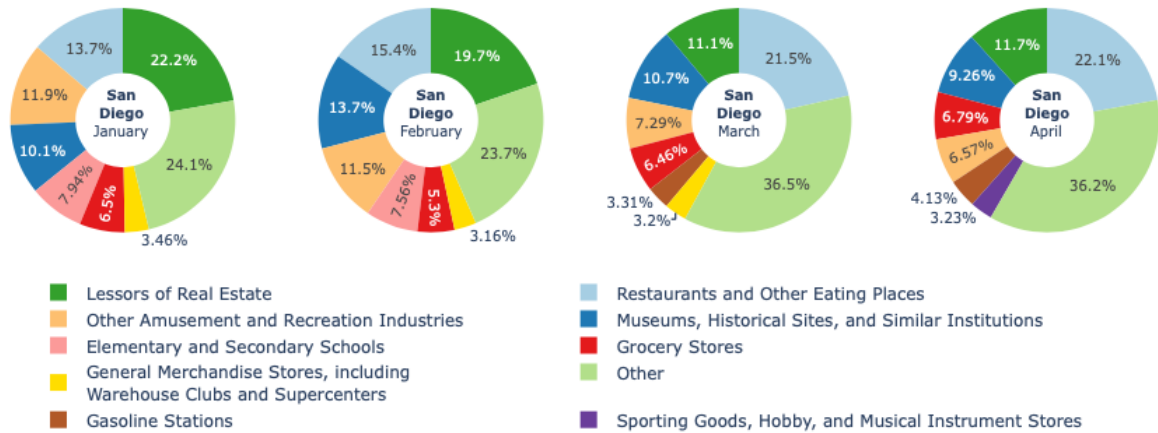

**Figure S12.** San Diego.

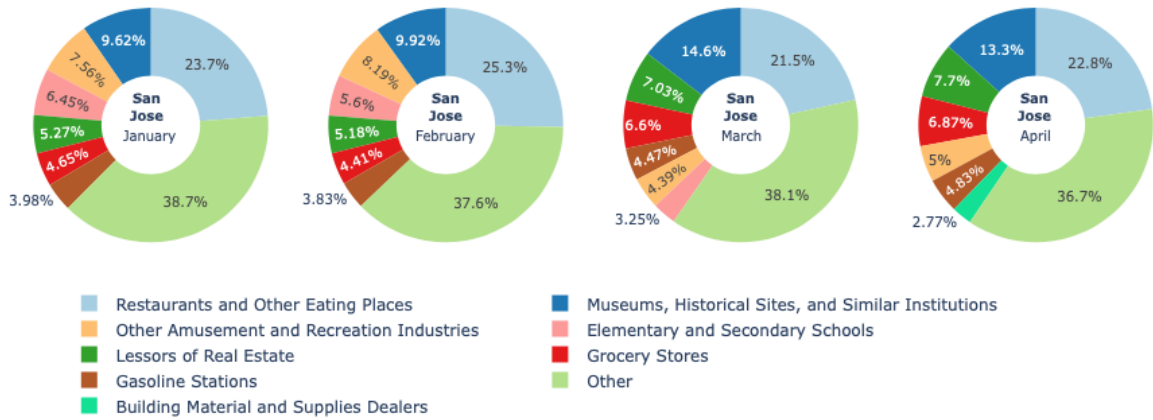

**Figure S13.** San Jose

## 4. Highly affected POIs in four studied weeks

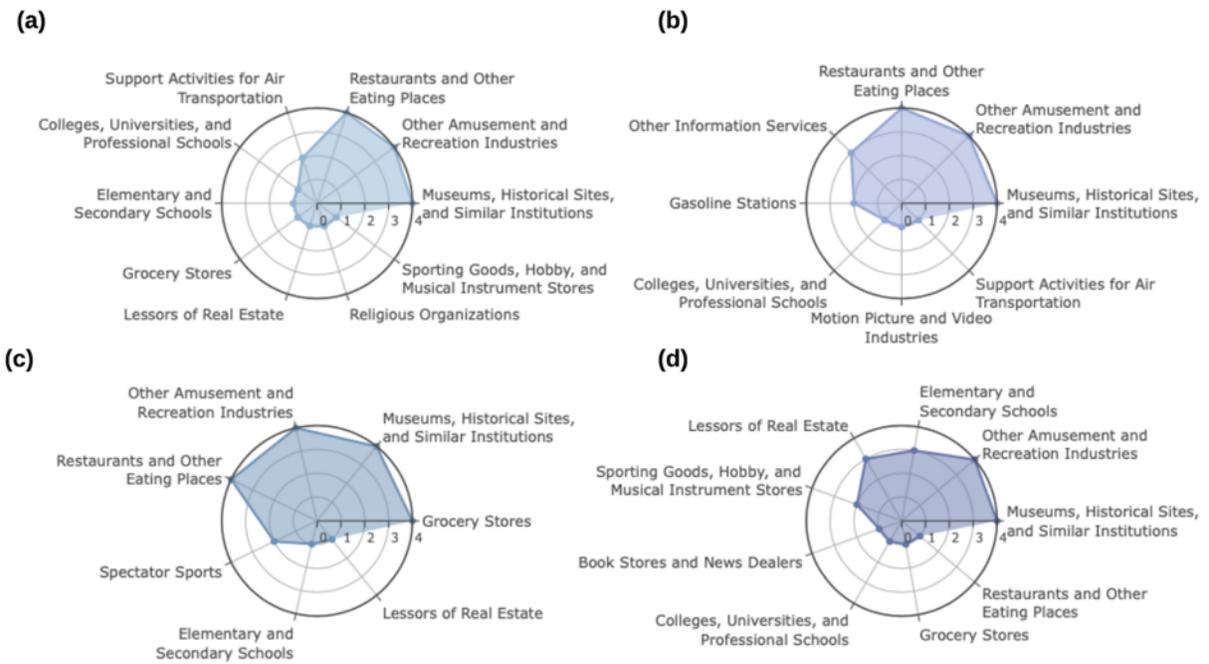

**Figure S14.** (a) Chicago, (b) Los Angeles, (c) Philadelphia, (d) San Jose.

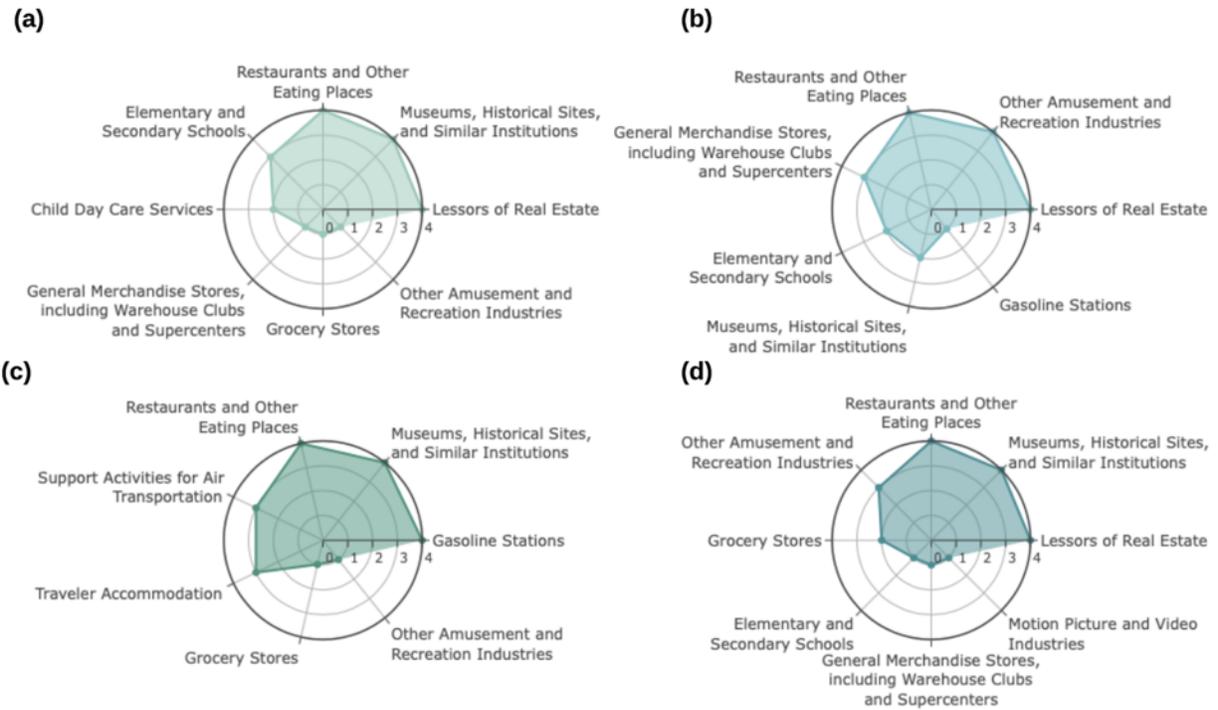

**Figure S15.** (a) Houston, (b) San Antonio, (c) Detroit, (d) Jacksonville.

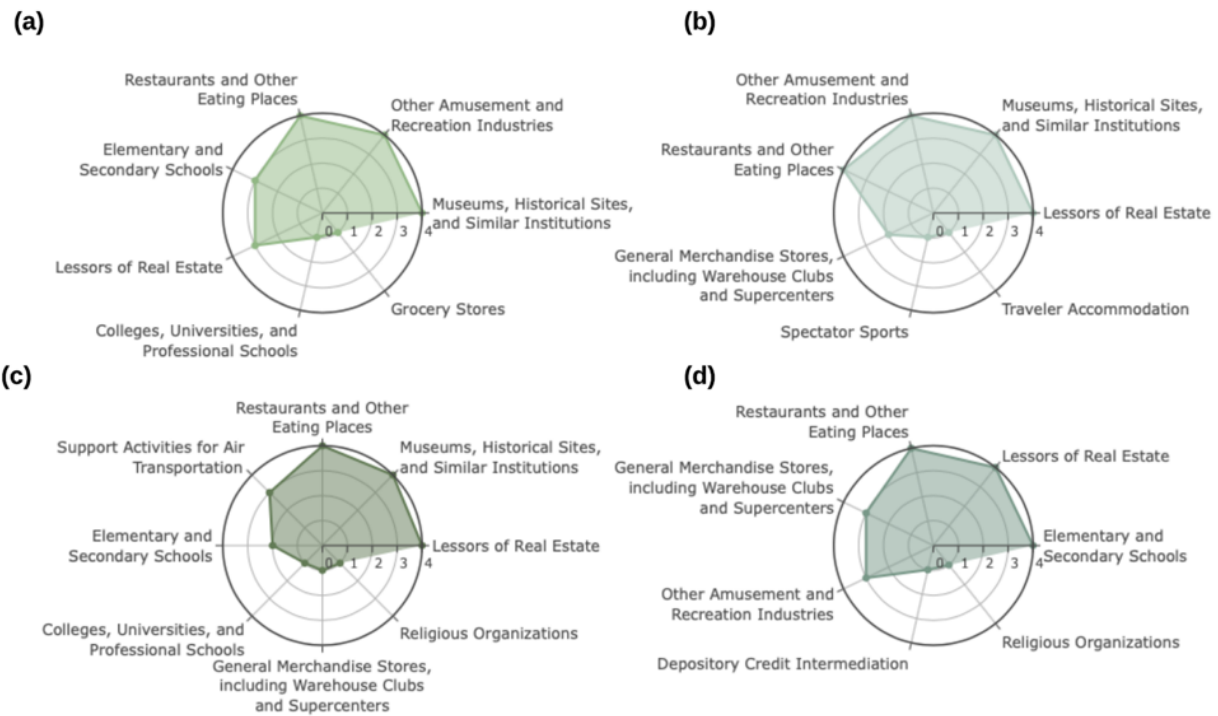

**Figure S16.** (a) San Diego, (b) Phoenix, (c) Dallas, (d) Fort Worth.

*4. Original data of four types of movements in 16 cities:*

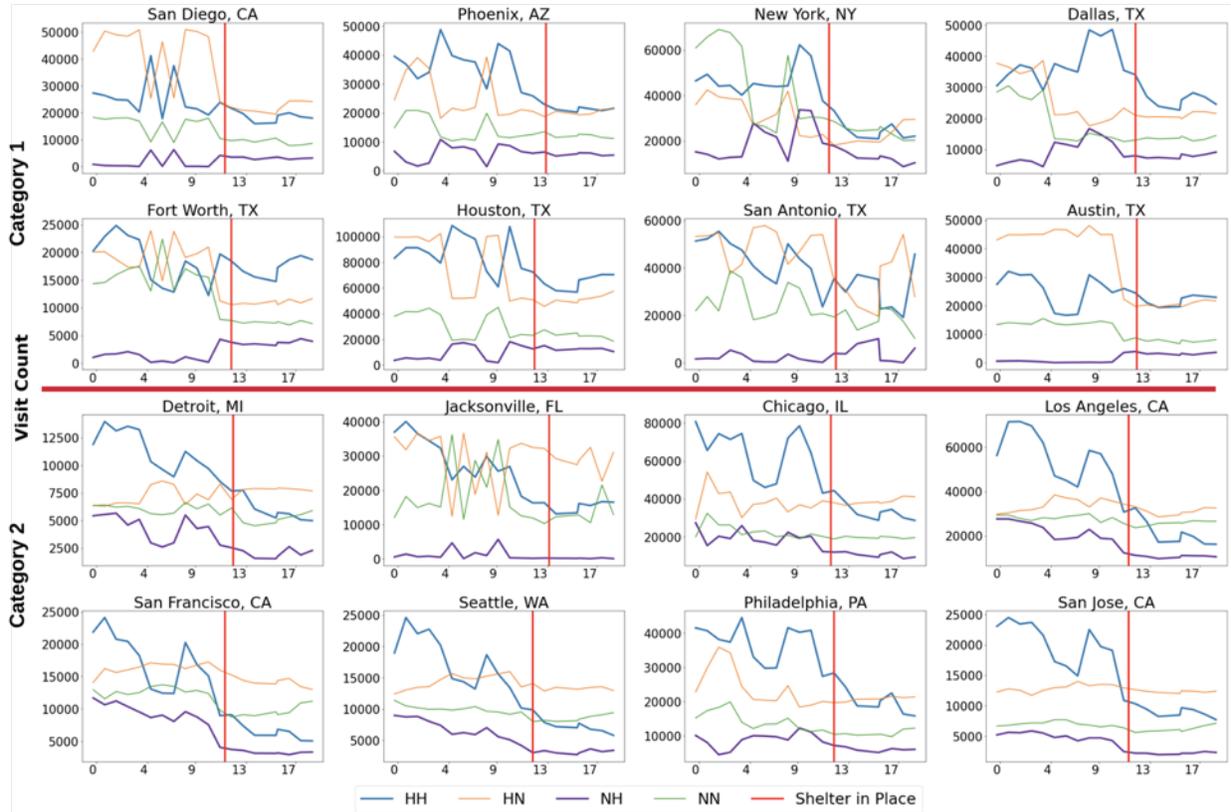

**Figure S20.** Absolute visits of four types of movements

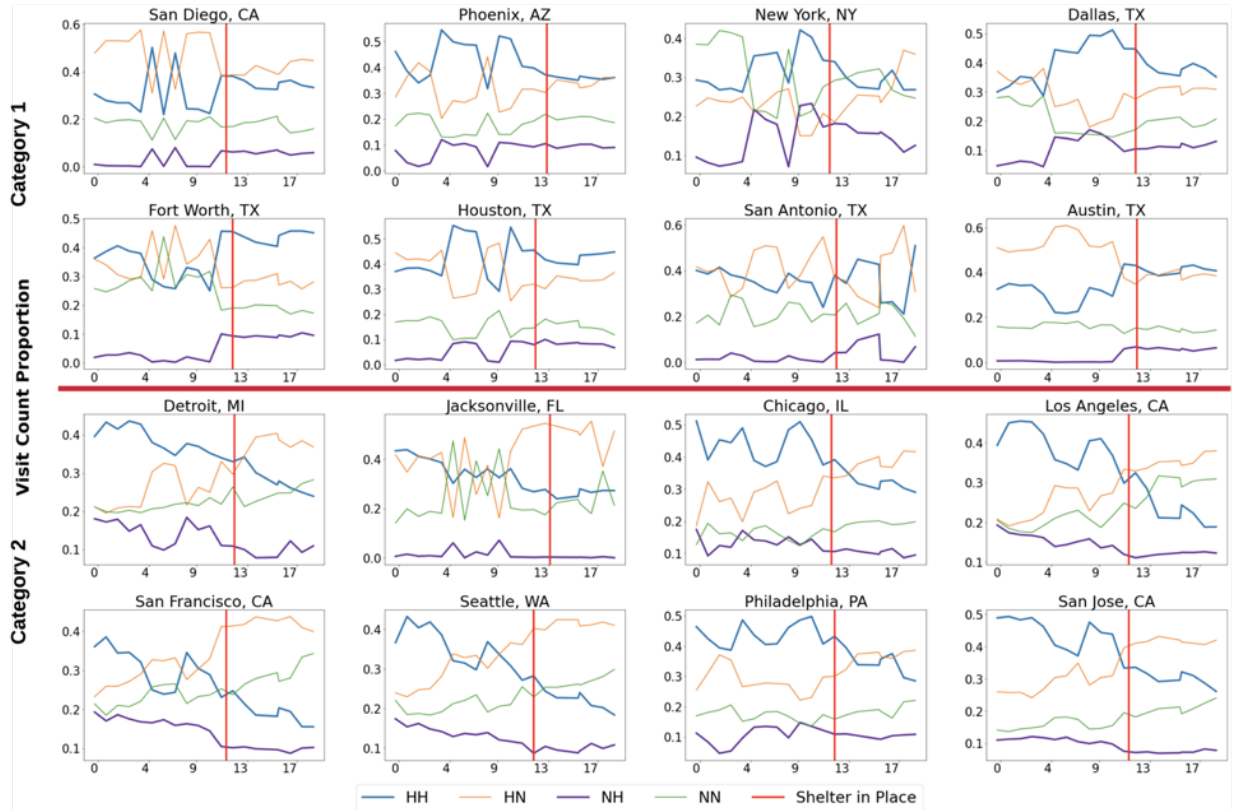

**Figure S21.** Proportion of four types of movements in 16 cities.